\documentclass[submission,copyright,creativecommons]{eptcs}
\providecommand{\event}{PC 2016} % Name of the event you are submitting to
\usepackage{breakurl}             % Not needed if you use pdflatex only.

\usepackage{graphicx}
\usepackage{color}

\usepackage[sort]{cite}
\usepackage{amssymb}  
\usepackage{amsthm}
\usepackage{amsmath}

\newcommand{\ket}[1]{|#1 \rangle}

\newcommand{\sandwich}[3]{\left \langle #1 \left| #2 \right| #3 \right\rangle}

\usepackage{hyperref}

\title{An Overview of Approaches to Modernize Quantum Annealing Using Local Searches}
\author{Nicholas Chancellor
\institute{Department of Physics, Durham University\\ South Road, Durham, UK}
\email{nicholas.chancellor@durham.ac.uk}
}
\def\titlerunning{An Overview of Approaches to Modernize Quantum Annealing Using Local Searches}
\def\authorrunning{N. Chancellor}
\begin{document}
\maketitle

\begin{abstract}
I describe how real quantum annealers may be used to perform local (in state space) searches around specified states, rather than the global searches traditionally implemented in the quantum annealing algorithm. The quantum annealing algorithm is an analogue of simulated annealing, a classical numerical technique which is now obsolete. Hence, I explore strategies to use an annealer in a way which takes advantage of modern classical optimization algorithms, and additionally should be less sensitive to problem mis-specification then the traditional quantum annealing algorithm.% Furthermore, I discuss how sequential calls to quantum annealers can be used to construct analogues of population annealing and parallel tempering which use quantum searches as subroutines. %The techniques given here can be applied not only to optimization, but also to sampling. %I examine the feasibility of these protocols on real devices and note that implementing such protocols should require minimal if any change to the current design of the flux qubit-based annealers by D-Wave Systems Inc.
\end{abstract}

\section{Quantum Annealing Algorithm vs. Local Search}

Recently, there has been much interest in using the Quantum Annealing
Algorithm (QAA) \cite{Finella1994,Farhi2001,Brooke1999} which utilizes
quantum tunnelling to aid in solving commercially interesting problems.
A complete list of all potential applications would be too long to
give here. However, applications have been studied in such diverse
fields as finance \cite{Marzec}, computer science \cite{Choi2010},
machine learning \cite{Adachi2015,Amin2016}, communications
\cite{Chancellor2015,Otsubo(2012),Otsubo(2014),Jordan(2006)}, graph
theory \cite{Vinci(2014)}, and aeronautics \cite{Coxson(2014)},
illustrating the importance of such algorithms to real world problems.
%While some of these applications rely on the ability of the QAA to
%perform optimization by finding the lowest energy state of a classical
%problem Hamiltonian, others such as \cite{Adachi2015,Denil,Rose(2014),Amin2016,Chancellor2015,Jordan(2006)},
%instead rely on the fact that open quantum systems effects allow for
%sampling of an approximate Boltzmann distribution. I will discuss
%both of these techniques in due course. 

The archetypal model for quantum annealing, because of its connection
to condensed matter physics as well as the fact that it can be implemented
on real devices is the transverse field Ising model,
with Hamiltonian $H(s)$ given by
\begin{equation}
H(s)=-A(s)\,\sum_{i}\sigma_{i}^{x}+B(s)\,H_{Problem}, \, \, \, \,\,\,\,\, H_{Problem}=-\sum_{i}h_{i}\sigma_{i}^{z}-\sum_{i,j\in\chi}J_{ij}\sigma_{i}^{z}\sigma_{j}^{z}.\label{eq:AnnealerHam}
\end{equation}

$H_{Problem}$ encodes the problem of interest, $\chi$ is the hardware graph, and
$A(s)$ and $B(s)$ are the annealing schedule, which determines how
the energy scales of the transverse and longitudinal terms change
with the annealing parameter, $s\in\left[0,1\right]$. The problem
is encoded by specifying the values of $h_{i}$ and $J_{ij}$. For
the QAA, $A(0)\gg B(0)$ and $A(1)\ll B(1)$, and $A(s)$ decreases
monotonically while $B(s)$ increases monotonically with increasing
$s$. Applying the QAA consists of monotonically increasing $s$ with
time such that the ground state of the system changes over time between
the (known) ground state of the transverse part of the Hamiltonian
(\,$\sum_{i}\sigma_{i}^{x}$\,) to the solution of the (classical) problem
to be solved, Eq.~(\ref{eq:AnnealerHam}). The search space of the transverse
Ising model is a hypercube where each vertex corresponds to a bitstring, the dimension is equal to the number of qubits, and the Hamming
distance between classical states corresponds to the number of edges
which must be traversed between the states. This structure is independent
of the interaction graph defined by $J_{ij}$ which, along with $h_{i}$,
determine the energy at each vertex.

I choose to focus on the transverse field Ising model for concreteness,
and because the action of the transverse field is a quantum analogue
of single bit flip updates in classical Monte Carlo methods. However, the
arguments presented in this paper should hold for most other search
spaces as well, with the Hamming distance replaced with a more general
notion of search space distance. Because the relevant effect of problem 
mis-specification is the energy difference in the states which are searched,
local searches can remain valid even if the global space is corrupted by a mis-specification.

The QAA can be though of as analogous to classical Simulated Annealing
(SA) in which quantum fluctuations mediated by the addition of non-commuting
terms to a classical Hamiltonian, play the role which temperature
plays in SA. Simple SA, however, has been superseded by more sophisticated
algorithms, such as parallel tempering~\cite{Swendsen1986,Earl2005},
population annealing~\cite{Hukushima2003,Matcha2010,Wang2015}, and
isoenergetic cluster updates~\cite{Zhu2015} to name a few. This then
begs the question of whether quantum annealing hardware can be used
in a clever way to gain the advantages of these modern classical algorithms,
by using a hybrid algorithm employing both quantum and classical search
techniques, or by using multiple quantum searches in a sequential
way to make algorithmic gains. 

The QAA, as it is currently designed, is not amenable to such adaptations.
It is a global search, and there is no obvious way to insert information,
from either a classical algorithm or previous runs of the QAA, in
a meaningful way to improve the performance. Furthermore, the QAA
is fundamentally different from classical annealing in that, due to
the famous no-cloning theorem~\cite{Wooters(1982)} of quantum mechanics,
we cannot determine exactly what the intermediate state of the system
is part way though the anneal. This is in direct contrast to SA, where
every intermediate state is known, and can be manipulated arbitrarily
to build better algorithms. For example, classical gains can be made
by running many runs in parallel and probabilistically replacing poor
performing copies with those which are performing well (population
annealing), or raising the temperature for those which perform poorly
and lowering it for those which perform well (parallel tempering). 

In order to build quantum versions, let us consider a subroutine similar
to QAA, but which performs a local search of a region of phase space
with a controllable size around a user selected initial state. The
input and output of a single step of this algorithm is completely
classical, so the no-cloning theorem is no longer a barrier and these
local quantum searches can be combined arbitrarily with both other
quantum searches and classical searches. 

Moreover, an effective temperature which can be used to construct analogues to 
parallel tempering and population annealing.  To construct this we first diagonalize
the Hamiltonian of a single qubit under quantum annealing $H_{1}(s')=-A(s')\,\sigma^{x}+B(s')\,\sigma^{z}$
to obtain the ground state ratio of probability amplitudes
$$\frac{\psi(1)}{\psi(2)}=\frac{\sqrt{A(s')^{2}+B(s')^{2}}}{A(s')}+\frac{B(s')}{A(s')}.$$
From this ratio an effective temperature can be derived by comparing to a Boltzmann distribution,
$$T_{eff}(s')=2\,\left[\ln\left(\left|\frac{\psi(1)}{\psi(2)}\right|^{2}\right)\right]^{-1}.$$ For details of these algorithms as well as more discussion of tolerance to problem mis-specification, use in thermal sampling, and feasibility in real devices, see \cite{journal_version}.

\section{Local Search on an Annealer\label{sec:Local_search}}

For a useful local search we desire two properties, firstly the search
should be local in the sense that it only explores a fraction of the
states in the state space and secondly the search should seek out
more optimal (lower energy) solutions over less optimal ones. Consider
a protocol to search the phase space near a chosen classical state
in the presence of a low temperature bath. The system is first initialized
at $s=1$ in a state which specifies the starting point of the algorithm
and therefore the region to be searched. Local search with a controllable
range is then performed by decreasing the annealing parameter $s$
in Eq.~\eqref{eq:AnnealerHam} to a prescribed value $s'$ (thereby
turning on a transverse field), possibly waiting for a period of time,
and then returning to $s=1$ and reading out the final state normally.
The low temperature bath will moderate transitions between states,
with detailed balance acting as a guarantee that more optimal states
will be favoured in the search. 

One model which has been able to successfully predict experimental
results \cite{Boixo2016,Dickson2013,Jonson2011,Chancellor2016}
is to assume decoherence acts in the energy eigenbasis. In this model,
which arises from a perturbative expansion in coupling strength \cite{Breuer_book,Albash2014},
coherence can be lost rapidly between energy eigenstates and transitions
between these states can be mediated by the bath but the eigenstates
themselves are not disrupted by the bath. Because the eigenstates
themselves will generally be highly quantum objects, even a completely
incoherent superposition of them can still support quantum effects.

Solving problems using tunnelling mediated by open quantum system effects
means that even if the system is initialized in an excited state,
interactions with the environment will cause probability transitions
to other eigenstates. Detailed balance implies that for a bath with
finite temperature the transitions will occur preferentially toward
lower energy states. Furthermore, if $A(s)$ is appropriately small
compared to $B(s)$ in Eq.~(\ref{eq:AnnealerHam}) then the quantum
fluctuations can be viewed as local fluctuations around a classical
state, the stronger $A(s)$ is compared to $B(s)$, the less local
this search will be. Consider the perturbative expansion around a
(non-degenerate) classical state $\ket{C(s=0)}$ which can be written
as,

\begin{equation}
\ket{C(s)}=\frac{1}{\mathcal{N}}\sum_{n=0}^{\infty}\left(\frac{A(s)}{B(s)}\right)^{n}\mathbf{D}_{n}\,\big(\,\sum_{i}\sigma_{i}^{x}\,\big)^{n}\ket{C(0)}\label{eq:class_pert}
\end{equation}
where $\mathbf{D}_{n}$ is a diagonal matrix which depends on the
spectrum of $H_{Problem}$ and $\mathcal{N}$ is a normalization factor.
If we assume dephasing noise, then the tunnelling rate between two
perturbed classical states, $\ket{C(s)}$ and $\ket{C'(s)}$ will
be proportional to $\sandwich{C(s)}{\sum_{i}\sigma_{i}^{z}}{C'(s)}$.
By inserting the state given in Eq.~\eqref{eq:class_pert}, we see that
\begin{equation}
\sandwich{C(s)}{\sum_{i}\sigma_{i}^{z}}{C'(s)}\propto\left(\frac{A(s)}{B(s)}\right)^{\mathcal{H}(C(0),\,C'(0))}+\cdots
\end{equation}
where $\mathcal{H}(C(0),C'(0))$ is the Hamming distance (number of
edges required to traverse on the hypercube) between the two classical
states and $\ldots$ indicates higher powers of $\frac{A(s)}{B(s)}$.
For small $\frac{A(s)}{B(s)}$, tunnelling between perturbed classical
states is therefore \emph{exponentially} suppressed in the Hamming
distance between the states. As an eigenstate of a transverse field
Ising model $\ket{C(s\neq0)}$ is a fundamentally quantum object which
will exhibit quantum entanglement and therefore will be able to mediate
tunnelling between classical states quantum mechanically. We therefore
expect a quantum advantage to be preserved within the local search.
By using quantum searches only locally we have removed the possibility
of gaining a quantum advantage for long range searches beyond the
range of each local search. However, for this price we gain a major
advantage, the classical long range search can be done using state-of-the-art
techniques such as parallel tempering or population annealing, therefore
a small quantum advantage in the local search still results in an
improvement over the underlying classical algorithm. By contrast,
traditional quantum annealing only represents algorithmic improvement
over classical methods if the quantum advantage is \emph{at least}
as large as the advantage which state-of-the-art classical techniques
such as parallel tempering have over simulated annealing.

The question is now how one could program the initial state. The initial
states required for the local search protocol are completely classical,
and therefore could be programmed directly by manipulating the qubits
in a classical way. Another completely classical method would be to
prepare a simple energy landscape where the desired state has the
lowest energy and first heat and than cool the system, thus preparing
it by classical thermal annealing. Both of these methods would require
additional controls or degrees of freedom which may not be accessible
on a real device. For this reason I will instead focus on preparing
the initial state using the standard QAA, which an annealer is able
to perform \emph{by definition}. This is accomplished by running the
QAA with a simple Hamiltonian to guarantee that the system is initialized
in a desired state $y$ (\,$y(i)\in\left\{ -1,1\right\}$\,) with a high
probability, for example $
H_{init}(y)=-\sum_{i}\,y(i)\,\sigma_{i}^{z}-\sum_{i,j\in\chi}y(i)\,y(j)\,\sigma_{i}^{z}\sigma_{j}^{z} $,
which is a gauge transform of an unfrustrated ferromagnetic system
in a strong field, and will have a very simple energy landscape and
a relatively large energy difference between the lowest energy and
first excited state. Annealing runs with this Hamiltonian therefore
should therefore reach the target state $y$ with a high probability.
After this step, one needs to be able to reprogram $H_{Problem}$
in Eq.~\eqref{eq:AnnealerHam} to be the Hamiltonian of the problem
in which we are interested. %I will discuss the feasibility of performing
%such a protocol on real annealers in Sec. \ref{sec:Hardware-Implementation}.

\begin{figure}
\begin{centering}
\includegraphics[width=.5\textwidth]{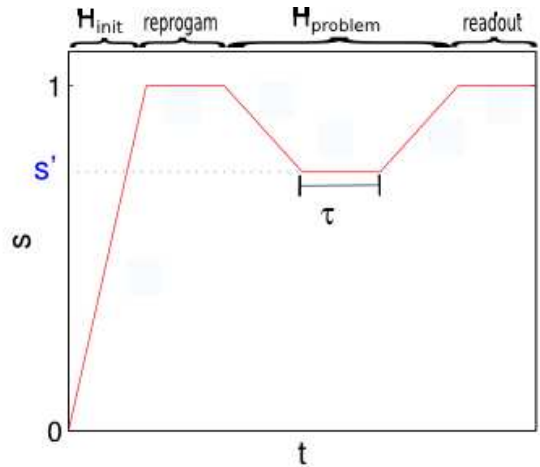}
\par\end{centering}

\caption{\label{fig:runback}Schematic representation of a single annealing
cycle to perform a local search. First the standard QAA is implemented
with the problem Hamiltonian given in Eq.~(\ref{eq:AnnealerHam}) to initialize
the qubits in the desired state. After this the Hamiltonian is reprogrammed
and the device is annealed to $s'$ and optionally allowed to remain
at that point for a time $\tau$. The device is then annealed to $s=1$
and read out. }

\end{figure}

Once we have the desired initial state and problem Hamiltonian programmed,
we simply need to turn on a desired strength of transverse field,
controlled by the value $s'$ shown in Fig. \ref{fig:runback}. It
may also be desirable to wait for a time $\tau$ before turning the
field off again and reading out the state. The readout does not need
to be any different than what is used with the standard QAA.
 It is worth pointing out that in the limit
each run is effectively a noisy continuous time quantum random walk
with a localized starting condition and a finite temperature bath.
In a related work I will examine the relationship between QAA and
quantum random walks in the context of global rather than local searches
\cite{Chancellor_in_prep}.

%The function ANNEALER\_CALL in algorithms \ref{alg:adaptive_s}, \ref{alg:parallel_tempering},
%and \ref{alg:population_annealing} can be constructed by repeatedly
%performing the annealing cycle protocol illustrated in Fig. \ref{fig:runback}
%with the same value of $s'$ and the same initial state $\ket{C(s=0)}$
%each time. This function than returns a list of the final state found
%in each successive annealing cycle, which is can be thought of as
%the results of a probabilistic local quantum search around $\ket{C(s=0)}.$

\section*{Acknowledgments}

The author was supported by EPSRC (grant ref: EP/L022303/1), and
thanks Viv Kendon for several critical readings of \cite{journal_version}
and useful discussions. The author further thanks Trevor Lanting,
Helmut Katzgraber, Gabriel Aeppli, Andrew G. Green, and Paul A. Warburton
for useful discussions.

%\nocite{*}
%\bibliographystyle{unsrt}
%\bibliographystyle{eptcs}
%\bibliography{Chancellor2016}

\end{document}